\documentclass[12pt]{iopart}

\usepackage{graphicx} 
  
\begin{document}

\title{Recent PHENIX Results on Open Heavy Flavor}
\author{J. Matthew Durham (for the PHENIX Collaboration\footnote{A list of members of the PHENIX Collaboration can be found at the end of this issue.}) }
\address{Department of Physics and Astronomy\\
Stony Brook University, Stony Brook, NY 11790, USA}
\ead{durham@skipper.physics.sunysb.edu}
\begin{abstract}

Throughout the history of the RHIC physics program, questions concerning the dynamics of heavy quarks have generated much experimental and theoretical investigation. A major focus of the PHENIX experiment is the measurement of these quarks through their semi-leptonic decay channels at mid and forward rapidity. Heavy quark measurements in $p+p$ collisions give information on the production of heavy flavor, without complications from medium effects. New measurements in $d+$Au and Cu+Cu indicate surprising cold nuclear matter effects on these quarks at midrapidity, and provide a new baseline for interpretation of the observed suppression in Au+Au collisions.  When considered all together, these measurements present a detailed study of nuclear matter across a wide range of system size and temperature. Here we present preliminary PHENIX measurements of non-photonic electron spectra and their centrality dependence in $d$+Au and Cu+Cu, and discuss their implications on the current understanding of parton energy loss in the nuclear medium.
\end{abstract}

\section{Introduction}

Charm and bottom quarks are dominantly produced by gluon fusion in the early stages of nuclear collisions, and thus probe the complete evolution of the medium.  The production baseline measured in $p+p$ collisions can be described by fixed order plus next to leading log perturbative QCD calculations within uncertainties \cite{PPG065}.  The large mass of heavy quarks ($M_{c,b} \approx $ 1.3 and 4.3 GeV/$c^{2}$, repectively) is expected to cause a suppression of gluon radiation at forward angles, known as the ``dead cone effect" \cite{deadcone}.  If radiative energy loss in the hot medium is the dominant suppression mechanism in Au+Au collisions, this leads to the expectation that $R_{AA}^{u,d} < R_{AA}^{c} < R_{AA}^{b}$.  In central Au+Au collisions, a large suppression has been measured relative to the yield in $p+p$ scaled by the number of nucleon-nucleon collisions, suggesting a significant energy loss by heavy quarks in the medium \cite{PPG066}.

The PHENIX collaboration measures open heavy flavor through the semi-leptonic decay of these quarks to electrons at midrapidity.  A cocktail of background electrons from hadronic decays and photon conversions is subtracted from the inclusive electron yield to isolate the contribution from heavy quarks (see \cite{PPG077} for a detailed description).  Preliminary PHENIX results using improved electron identification techniques have extended the range of this measurement in $p+p$ collisions from $p_{T}$ = 9 GeV/$c$ out to 14 GeV/$c$, and agree well with previously published data.  

\begin{figure}[h]
	\centering
		\includegraphics[width=0.45\textwidth]{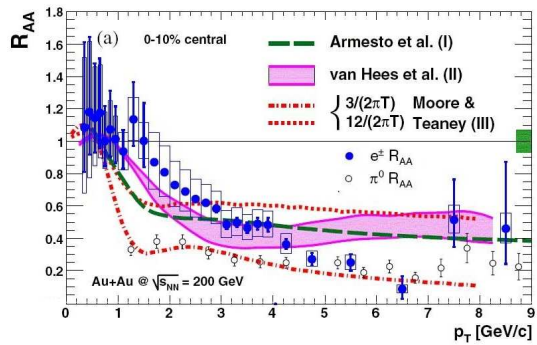}
	\caption{The nuclear modification factor $R_{AA}$ for electrons from heavy flavor decays and neutral pions \cite{PPG066}.}
	\label{fig:figure1}
\end{figure}

This technique of measuring heavy quarks through their decay to electrons does not allow separation of the contributions from charm quarks and bottom quarks.  However, perturbative QCD calculations and measurements in $p+p$ collisions at RHIC show that electrons from bottom decays begin to dominate over those from charm at $p_{T} \approx 5$GeV/$c$ \cite{PPG065_pQCD1}, \cite{PPG094}, \cite{cbSTAR}.  The ratio of electrons from charm to those from bottom measured in $p+p$ would hold in Au+Au collisions only if the charm and bottom quark energy loss were identical.  In the conventional picture of quark energy loss, charm loses more energy than bottom, which would shift the electrons from charm decay to lower transverse momentum and thereby increase the concentration of electrons from bottom at high $p_{T}$, relative to the $p+p$ baseline.  In any case, the heavy flavor electron $R_{AA}$ shown in Fig. \ref{fig:figure1} seems to suggest not only that bottom is highly suppressed, but that it is suppressed more than charm.  This is, in fact, the exact opposite of expectations from the dead cone effect.

\section{Cold Nuclear Matter Effects}

It is not immeditely clear if the suppression observed in Fig. \ref{fig:figure1} can be solely contributed to energy loss in the hot medium.  The heavy quark production mechanism is sensitive to the gluon density in the nucleus, so inital state effects (such as shadowing or gluon saturation) could possibly affect charm and bottom cross sections.  Also, the mass-ordered Cronin enhancement that is observed for light-flavored hadrons \cite{PPG030} could conceivably play a role in the heavier $D$ and $B$ meson families.  To examine these phenomena, it is necessary to study $d+$Au collisions, where nuclear effects can be measured without the complications from the hot medium.

\begin{figure}
	\centering
		\includegraphics[width=0.45\textwidth]{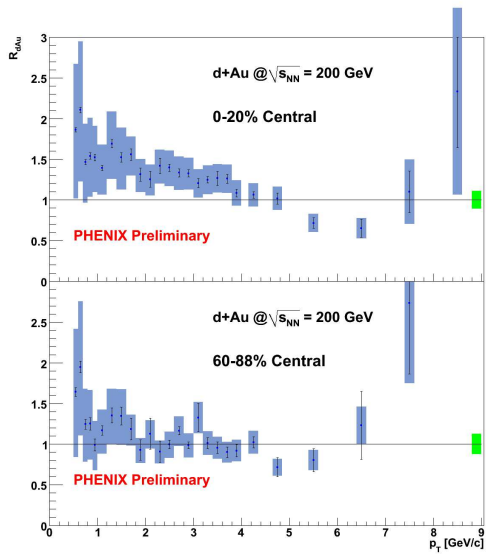}
	\caption{The nuclear modification factor $R_{dA}$ for electrons from heavy flavor decays.}
	\label{fig:figure2}
\end{figure}

Fig. \ref{fig:figure2} shows the nuclear modification factor $R_{dA}$ of electrons from heavy flavor produced in $d+$Au collisions during RHIC's Run-8.  In the most peripheral $d+$Au collisions, the heavy flavor production agrees with the scaled $p+p$ result, within uncertainties.  For the most central $d+$Au collisions, there is evidence of cold nuclear matter effects at moderate $p_{T}$, although the large uncertainties inherent to the cocktail method of background subtraction severely limit the precision of the measurement.  The large suppression seen in central Au+Au events is not observed, suggesting that the effect is indeed due to energy loss in the hot nuclear medium.  On the contrary, a possible enhancement is present for $p_{T} <$ 4 GeV/$c$, although the large uncertainties limit this observation.  

It is interesting to note that this is the same $p_{T}$ range that electrons from charm decays are expected to dominate, and the possible enhancement disappears as the relative proportion of electrons from bottom decays increases.  These cold nuclear matter effects are expected to be present in Au+Au collisions, but are convoluted with suppression effects in hot nuclear matter.  If there is some CNM enhancement of open charm but not bottom, this could potentially explain the heavy quark energy loss paradox apparent in Fig. \ref{fig:figure1}. Charm quarks could suffer a larger energy loss in the medium than bottom, and still appear to be less suppressed because of an enhanced initial state.  If all quark flavors have the same energy loss in the medium, the pion and bottom quark $R_{AA}$ would be equivalent (as can be inferred from the high $p_{T}$ behavior in Fig. \ref{fig:figure1}), while charm quarks with an enhanced initial state would have a larger $R_{AA}$.

\begin{figure}
	\centering
		\includegraphics[width=0.6\textwidth]{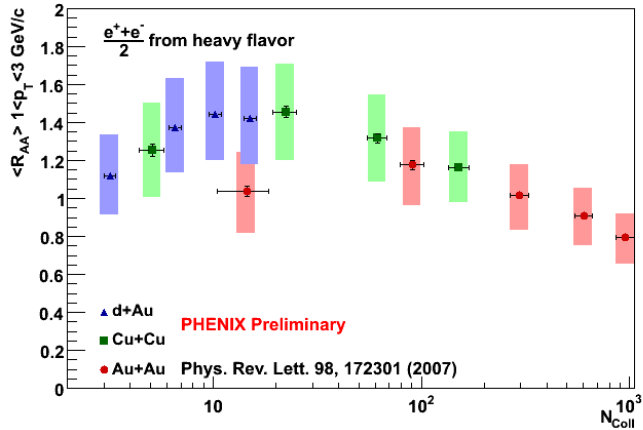}
	\caption{The average nuclear modification factor $R_{AA}$ for electrons from heavy flavor decays for 1 $< p_{T} <$ 3 GeV/c.}
	\label{fig:figure3}
\end{figure}

We turn to collisions of Cu nuclei to more precisely probe the range of $<N_{coll}>$ between central $d+$Au and Au+Au.  Using similar techniques, the heavy flavor electron yield is determined for 4 centrality bins in Cu+Cu, which cover up to $<N_{coll}> \approx$ 100.  Fig. \ref{fig:figure3} shows the average value of $R_{AA}$ for 1 $< p_{T} <$ 3 GeV/c (which is well within the charm-dominated portion of the electron spectra), for $d+$Au, Cu+Cu, and Au+Au collisions as a function of $<N_{coll}>$.  The horizontal error bars represent the uncertainty on $<N_{coll}>$ for each centrality.  Although subject to large systematic uncertainties, a possible enhancement is seen with increasing $<N_{coll}>$, that then disappears in central Au+Au collisions.

\section{Summary}

The large suppression of electrons from heavy flavor decays observed in central Au+Au collisions at $\sqrt{s_{NN}} = 200$ GeV is not observed in $d+$Au and peripheral Cu+Cu collisions.  This implies that the suppression is due to energy loss in the hot nuclear medium produced in central collisions of large nuclei, rather than initial state effects intrinsic to the nucleus or parton energy loss in cold nuclear matter. Forthcoming finalized results in $d+$Au and Cu+Cu with reduced uncertainties will allow more quantitative statements about possible cold nuclear matter effects on heavy quark production.

\end{document}